\documentclass[10pt, conference]{IEEEtran}
\usepackage[utf8]{inputenc}
\usepackage[detect-all]{siunitx}
\usepackage{array,tabularx,calc}
\usepackage{mathtools}
\usepackage{amsfonts}
\usepackage{graphicx}
\usepackage{amssymb}
\usepackage{amsfonts}
\usepackage{pbox}
\usepackage{url}
\usepackage[]{algorithm, algpseudocode}
\usepackage{tikz}

\DeclareSIUnit{\belmilliwatt}{Bm}
\DeclareSIUnit{\dBm}{\deci\belmilliwatt}
\DeclareSIUnit{\belisotropic}{Bi}
\DeclareSIUnit{\dBm}{\deci\belisotropic}
\DeclareSIUnit{\bit}{bit}

\ifCLASSOPTIONcompsoc	\usepackage[caption=false,font=normalsize,labelfont=sf,textfont=sf]{subfig}
\else					\usepackage[caption=false,font=footnotesize]{subfig}
\fi

\makeatletter
\usepackage[]{algorithm, algpseudocode}
\makeatletter
\renewcommand{\ALG@beginalgorithmic}{\small}

\usepackage[capitalise]{cleveref}	
    \crefformat{equation}{(#2#1#3)}	

\usepackage[noadjust]{cite}

\begin{document}

\usetikzlibrary{arrows}
\usetikzlibrary{shapes}
\newcommand{\mymk}[1]{%
	\tikz[baseline=(char.base)]\node[anchor=south west, draw,rectangle, rounded corners, inner sep=0.1pt, minimum size=3.5mm,
	text height=2mm](char){\ensuremath{#1}} ;}

\newcommand*\circled[1]{\tikz[baseline=(char.base)]{
	\node[shape=circle,draw,inner sep=0.1pt] (char) {#1};}}

\title{Traffic-aware Gateway Placement\\for High-capacity Flying Networks}

\author{\IEEEauthorblockN{André Coelho, Helder Fontes, Rui Campos, Manuel Ricardo}
	\IEEEauthorblockA{INESC TEC and Faculdade de Engenharia, Universidade do Porto, Portugal\\
		\{andre.f.coelho, helder.m.fontes, rui.l.campos, manuel.ricardo\}@inesctec.pt}}

\maketitle

\begin{abstract}
The ability to operate virtually anywhere and carry payload makes Unmanned Aerial Vehicles (UAVs) perfect platforms to carry communications nodes, including Wi-Fi Access Points (APs) and cellular Base Stations (BSs). This is paving the way to the deployment of flying networks that enable communications to ground users on demand. Still, flying networks impose significant challenges in order to meet the Quality of Experience expectations. State of the art works addressed these challenges, but have been focused on routing and the placement of the UAVs as APs and BSs serving the ground users, overlooking the backhaul network design. The main contribution of this paper is a centralized traffic-aware Gateway UAV Placement (GWP) algorithm for flying networks with controlled topology. GWP takes advantage of the knowledge of the offered traffic and the future topologies of the flying network to enable backhaul communications paths with high enough capacity. The performance achieved using the GWP algorithm is evaluated using ns-3 simulations. The obtained results demonstrate significant gains regarding aggregate throughput and delay.  
\end{abstract}


\section{Introduction}
\begin{figure}
	\setlength\abovecaptionskip{-0.1\baselineskip}
	\centering
	\includegraphics[width=1\linewidth]{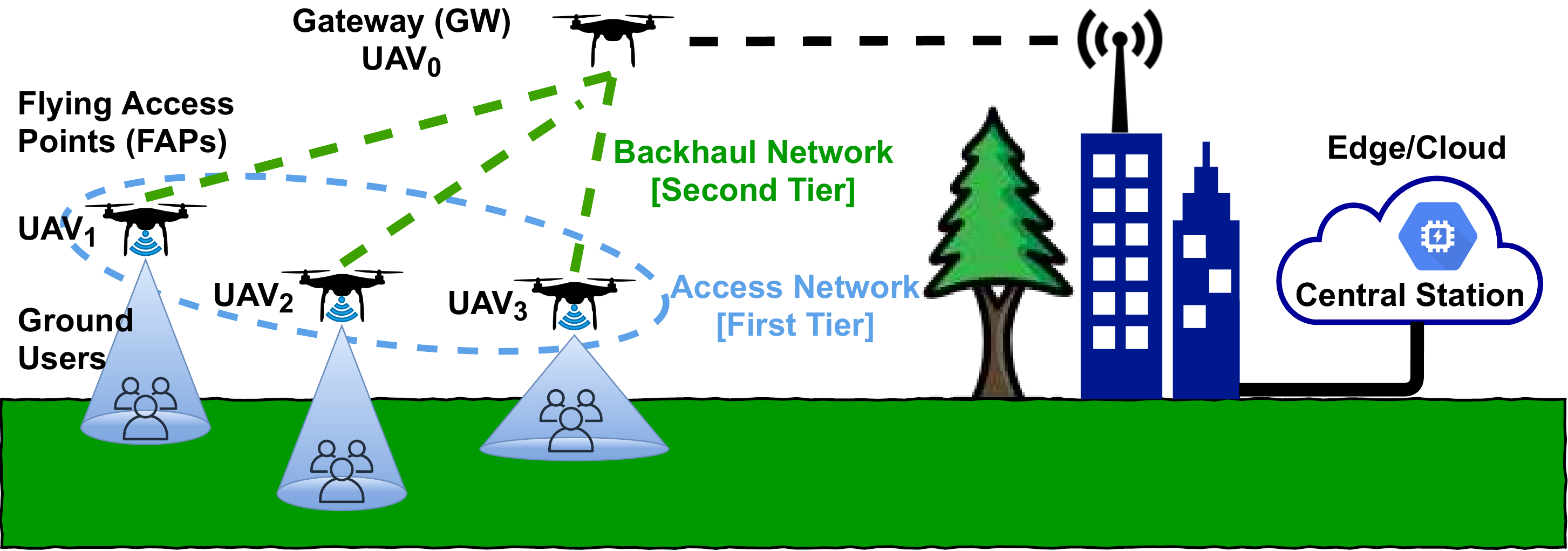}
	\caption{Multi-tier flying network providing Internet connectivity to ground users. The flying network is controlled by a
Central Station (CS) deployed on the Edge or in the Cloud.}
	\label{fig:conceptual-scenario}
\end{figure}

In the recent years the usage of Unmanned Aerial Vehicles (UAVs) has emerged to provide communications in areas without network infrastructure and to enhance the capacity of existing networks in temporary events~\cite{Almeida2018, Almeida2019, zhong2019}. The ability to operate virtually anywhere, as well as their hovering, mobility, and on-board payload capabilities make UAVs perfect platforms to carry communications nodes, including Wi-Fi Access Points (APs) and cellular Base Stations (BSs)~\cite{Hayat2016}. This is paving the way to the deployment of flying networks that enable communications to ground users anywhere, anytime. A reference example is the WISE project~\cite{WISE:online}, which proposes a novel communications solution based on Flying Access Points (FAPs) that position themselves according to the traffic demand of the ground users, as depicted in \cref{fig:conceptual-scenario}.\looseness=-1  

Still, flying networks impose significant routing challenges. Firstly, radio link disruptions may occur, due to the high-mobility of UAVs. On the other hand, there is inter-flow interference between neighboring UAVs that need to be close to each other to establish high capacity air-air radio links. These problems were addressed in~\cite{Coelho2018, Coelho2019}, where we have proposed the RedeFINE routing protocol and the Inter-flow Interference-aware Routing (I2R) metric, which constitute a centralized routing solution that defines in advance the forwarding tables and the instants they shall be updated in the UAVs, enabling uninterruptible communications.\looseness=-1 

When it comes to the UAV placement problem, state of the art works have been focused on the UAVs acting as APs and BSs, aiming at enhancing the radio coverage and the number of ground users served, as well as improving the Quality of Service (QoS) offered \cite{kalantari2017, alzenad2018}. Even though users are directly affected by the QoS and Quality of Experience (QoE) provided by the access network, the backhaul network, including the gateway (GW) placement, needs to be carefully designed in order to meet the variable traffic demand of the FAPs. This aspect has been overlooked in the state of the art.\looseness=-1  

The main contribution of this paper is a centralized traffic-aware GW UAV Placement (GWP) algorithm for flying networks with controlled topology. GWP takes advantage of both the knowledge of the offered traffic and the future topologies of the flying network to enable backhaul communications paths with high enough capacity to accommodate the traffic demand of the ground users. The performance achieved using GWP was evaluated using ns-3~\cite{ns-3}, allowing to demonstrate significant gains regarding aggregate throughput and delay.\looseness=-1 
 
The rest of the paper is organized as follows. 
\cref{sec:soa} presents the state of the art on GW placement approaches in wireless networks in general. 
\cref{sec:system_model} defines the system model. \cref{sec:problem_formulation} formulates the problem. \cref{sec:gw_placement} presents the GWP algorithm, including its rationale and a numerical analysis for a simple scenario. \cref{sec:performance_evaluation} addresses the performance evaluation, including the simulation setup, the simulation scenarios, the performance metrics, and the simulation results. Finally, \cref{sec:conclusions} points out the main conclusions and directions for future work.\looseness=-1 

\section{State of the Art~\label{sec:soa}}
In the literature, GW placement in wireless networks is a common problem. Over the years, different studies have been carried out~\cite{maolin2009gateways, seyedzadegan2013zero, targon2010joint, jahanshahi2019gateway}. However, the majority of them aim at minimizing the number of GWs while optimizing their placement, in order to meet some QoS metrics, including throughput and delay, and reducing the energy consumption. In~\cite{muthaiah2008single}, the authors show how the GW placement and the transmission power have a significant impact on the network throughput. For that purpose, they evaluate the performance of different heuristics. However, they do not consider the communications nodes' traffic demand. In~\cite{larsen2017}, the authors show how the placement of a UAV performing the role of network relay between ground nodes affects the communications' performance; nevertheless, the study considers only a pair of ground nodes. A model-free approach to determine the optimal position of a relay UAV is proposed in \cite{zhong2019}. Since it relies on real-time measurements, its main drawback is the convergence time required to achieve the optimal position.\looseness=-1    

An analogy between the GW placement and the sink placement in the context of Wireless Sensor Networks (WSNs) can be made, since both are in charge of receiving all the traffic generated within a network. In~\cite{poe2007minimizing}, the authors describe different sink placement approaches and explore their advantages and disadvantages. However, in WSNs the traffic demand is significantly lower and it is not generated continuously over time. In addition, the sink placement in this type of networks aims at minimizing the energy consumption or the delay, and increase the network lifetime, rather than enabling high-capacity paths, which is a key-requirement in flying networks providing Internet access.\looseness=-1

State of the art works have been focused on the placement of the UAVs as APs and BSs serving the ground users, overlooking the backhaul network design \cite{Almeida2018, Almeida2019, kalantari2017, alzenad2018}. Even though users are directly affected by the QoS and QoE provided by the access network, the backhaul network, including the GW placement, needs to be carefully designed in order to meet the variable traffic demand of the FAPs. This paper introduces a differentiating factor since it aims at defining in advance the position of a flying GW taking advantage of the controlled mobility over the communications nodes being served.\looseness=-1

\section{System Model~\label{sec:system_model}}
The flying network consists of $N$ UAVs that are controlled by a Central Station (CS). Two types of UAVs are assumed to compose the flying network, as depicted in \cref{fig:conceptual-scenario}: 1) FAPs, which provide Internet access to ground users; 2) a GW UAV, which connects the flying network to the Internet. The flying network is organized into a multi-tier architecture, as depicted in \cref{fig:conceptual-scenario}, being especially targeted at taking advantage of short range high-directional air-air radio links, which provide high bandwidth channels and low inter-flow interference. The CS can be deployed anywhere on the Edge or in the Cloud. The CS is in charge of periodically: 1) defining the updated positions of the FAPs by running the NetPlan algorithm~\cite{Almeida2018} or similar FAPs placement algorithm, so that the FAPs meet the traffic demand of the ground users, which is collected by the FAPs themselves and transmitted to the CS; 2) calculating the updated forwarding tables to be used by the FAPs, by running RedeFINE~\cite{Coelho2018} or any other state of the art routing approach; and 3) determining the updated GW UAV position, in order to enable links that accommodate the FAPs' traffic demand. The CS benefits from a holistic view of the network, including the updated positions and the traffic demand of the FAPs. This information is used to calculate in advance the forwarding tables and the position of the GW UAV. Finally, the CS sends the forwarding tables and the updated positions to both the FAPs and GW UAV, which configure and position themselves accordingly.\looseness=-1

\section{Problem Formulation~\label{sec:problem_formulation}}
In the following, we formulate the problem addressed in this paper. At time $t_k = k \times \Delta t, k \in \mathbb{N}_0$ and $\Delta t \in \mathbb{R}$, where $\Delta t \gg \SI{1}{\second}$ is the update period defined by the FAPs placement algorithm, the flying network is represented by a directed graph $G(t_k)=(V, E(t_k))$, where $V=\{0, ..., N-1\}$ is the set of UAVs $i$ positioned at $P_i=(x_i, y_i, z_i)$ inside a cuboid, $E(t_k) \subseteq  V \times V $ is the set of directional links between UAVs $i$ and $j$ at $t_k$, $i, j \in V$, and $(i,j)\in E(t_k)$. The wireless channel between two UAVs is modeled by the Free-space path loss model, since a strong Line of Sight (LoS) component dominates the links between UAVs flying dozens of meters above the ground.\looseness=-1 

Let us assume that $UAV_i$, $i\in \{1, ..., N-1\}$, performs the role of FAP and transmits a traffic flow of bitrate $T_i(t_k)$~\SI{}{bit/s} during time slot $t_k$ towards $UAV_0$, acting as GW UAV. In this case, we have a tree $T(V, E_T)$ that is a subgraph of $G$, where $E_T \subset E$ is the set of direct links between $UAV_i$ and $UAV_0$. This tree defines the flying network active topology. The flow $F_{0,i}$ with $T_i(t_k)$~\SI{}{bit/s} demands a minimum capacity $C_{0,i}(t_k)$, in bit/s, for the wireless link available from $UAV_0$ to $UAV_i$ at time $t_k$. $F_{0,i}$ is received at $UAV_0$ from $UAV_i$ with bitrate $R_i(t_k)$ bit/s. The maximum channel capacity is equal to $C^{MAX}$ bit/s. The wireless medium is shared and we assume that every $UAV_i$ can listen to any other $UAV_j$, including $UAV_0$. For this reason, the Carrier Sense Multiple Access with Collision Avoidance (CSMA/CA) mechanism is employed for Medium Access Control (MAC), which is in charge of avoiding collisions of network packets by enabling transmissions only when the channel is sensed to be idle.\looseness=-1

Considering $C_{0,i}(t_k)$ as the capacity of the bidirectional wireless link between $UAV_0$ and $UAV_i$ at time $t_k$, and $N-1$ UAVs generating a traffic flow $F_{0,i}$ with bitrate $T_i(t_k)$ bit/s towards $UAV_0$, we aim at determining at any time instant $t_k$ the position of $UAV_0$, $P_0 = (x_0, y_0, z_0)$, and the transmission power $P_T$ of the UAVs, considering $P_T^{MAX}$ as the maximum transmission power allowed for the wireless technology used, so that $C_{0,i}(t_k)$ is high enough to accommodate $T_i(t_k)$ bit/s, while the overall network capacity, $C(t_k) = \sum_{i=1}^{N-1}C_{0,i}(t_k)$, is minimized. By minimizing $C(t_k)$, a lower transmission power is required, which in turn allows to decrease the interference between the communications nodes and the energy consumption, improving the overall network performance. Our objective function is defined in \cref{eq:objective-function}. Since the wireless channel is shared by multiple communications nodes, the actual capacity of the wireless links is difficult to characterize mathematically. For this reason, in this paper, we propose a heuristic algorithm to solve the problem.\looseness=-1 

\begin{small}
\begin{subequations}
	\begin{alignat}{8}
		  & \!\underset{P_T, (x_0, y_0, z_0)}{\textrm{minimize}} &  & C(t_k)=\sum_{i=1}^{N-1}C_{0,i}(t_k)\label{eq:objective-function}\\
		  & \text{subject to:} & & 0 \leq P_T \leq P_T^{MAX}\label{eq:constraint1}\\
		  &     &     & C(t_k) \leq C^{MAX}\label{eq:constraint2}\\
		  &     &     & 0 < T_i(t_k) \leq C_{0,i}(t_k), i \in \{1,...,N-1\}\label{eq:constraint3}\\
		  &     &     & 0 \leq x_i \leq x^{MAX}, i \in \{0,...,N-1\}\label{eq:constraint4}\\
          &     &     & 0 \leq y_i \leq y^{MAX}, i \in \{0,...,N-1\}\label{eq:constraint5}\\	
          &     &     & 0 \leq z_i \leq z^{MAX}, i \in \{0,...,N-1\}\label{eq:constraint6}\\
		  &     &     & (0, i), (i, 0) \in E(t_k), i \in \{1, ..., N-1\} \label{eq:constraint7}\\
		  &     &     & (x_0, y_0, z_0) \neq (x_i, y_i, z_i), i \in\{1\smash{,}...\smash{,}N-1\}\label{eq:constraint8}
	\end{alignat}
\end{subequations}
\end{small}

\section{Traffic-aware Gateway UAV Placement~\label{sec:gw_placement} algorithm}
The traffic-aware Gateway UAV Placement (GWP) algorithm is presented in this section, including its rationale and a numerical analysis for a simple scenario.

\subsection{Rationale}
The GWP algorithm takes advantage of the centralized view of the flying network available at the CS, which is obtained as presented in Section \ref{sec:system_model}. For the sake of simplicity, we omit $t_k$ in what follows. Considering the future positions of $UAV_i$ and the bitrate of the traffic flow $F_{0,i}$, $T_i$, we aim at guaranteeing that a wireless link towards $UAV_0$ (GW UAV) has a minimum Signal-to-Noise Ratio (SNR), $SNR_{i}$, that enables the usage of a Modulation and Coding Scheme (MCS) index, $MCS_i$, capable of transmitting $T_i$~\SI{}{bit/s}. Conceptually, if $MCS_i$ is ensured, then the network will be able to accommodate $T_i$, which maximizes the amount of bits received in the GW UAV.\looseness=-1 

The selection of $MCS_i$ imposes a minimum $SNR_i$, considering a constant noise power $P_N$. Then, if the transmission power $P_{T}$ is known, we can calculate the maximum distance $d_{max_i}$ between $UAV_i$ and $UAV_0$, using the Free-space path loss model defined in \cref{eq:friis-propagation-model} in \SI{}{\decibel}, where $f_i$ is the carrier frequency and $c$ represents the speed of light in vacuum.\looseness=-1 

\begin{small}
\begin{equation}
	\begin{aligned}
	SNR_{i} \smash{=} P_{T} \smash{-} 20\log_{10}(d_{max_i}) \smash{-} 20\log_{10}(f_i) \smash{-} 20\log_{10} \bigl(\begin{smallmatrix}
	\frac{4\smash{\times}\pi}{c}
	\end{smallmatrix}\bigr) 
	\smash{-}P_N
	\end{aligned}
	\label{eq:friis-propagation-model}
\end{equation}
\end{small}

\begin{figure}
	\centering
	\setlength\abovecaptionskip{-0.1\baselineskip}
	\includegraphics[width=0.85\linewidth]{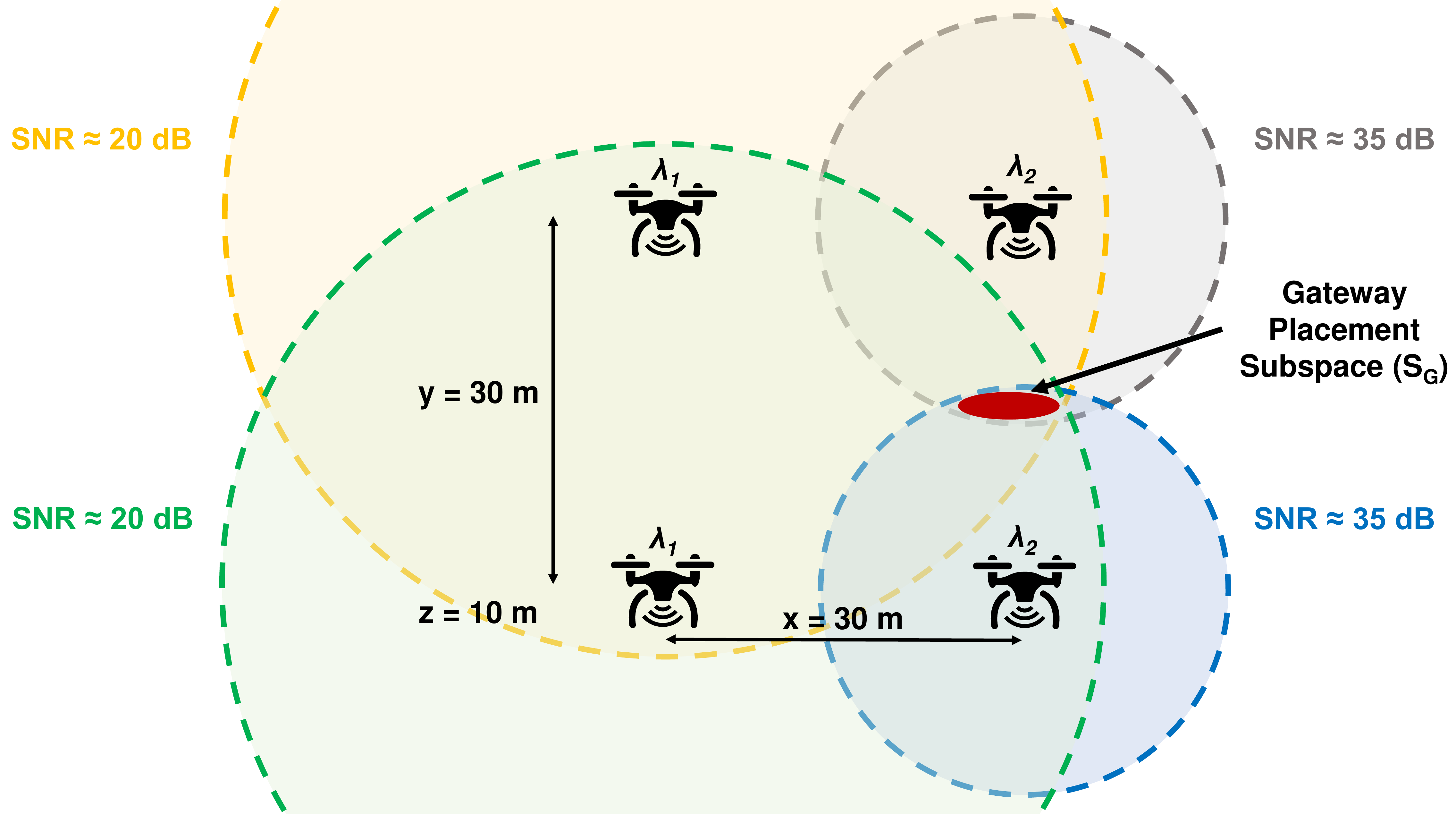}
	\caption{Gateway Placement Subspace ($S_G$) in a two-Dimensional (2D) space, which results from the intersection of the spheres, centered at each UAV, with radius equal to the maximum distance compliant with the target SNR values.}
	\label{fig:gw-acceptance-subspace}
\end{figure}

In the three-Dimensional (3D) space, $d_{max_i}$ corresponds to the radius of a sphere centered at $UAV_i$, inside which $UAV_0$ should be placed. Considering $N-1$ UAVs, the placement subspace for positioning $UAV_0$ is defined by the intersection of the corresponding spheres $i \in \{1, ..., N-1\}$; we refer to this subspace as the Gateway Placement Subspace, $S_G$, as depicted in \cref{fig:gw-acceptance-subspace}. In order to simplify the process of calculating $S_G$, we follow Algorithm~A, which iteratively allows obtaining the point $P_0 = (x_0, y_0, z_0)$ for positioning $UAV_0$ and the transmission power $P_T$ that we assume to be the same for all UAVs.\looseness=-1 

\begin{tiny}
\begin{algorithm}
	\renewcommand{\thealgorithm}{A}
	\caption{-- GWP Algorithm}
	\label{alg:gwp}
	\begin{algorithmic}[1]
		\State $P_T = 0$ \Comment{\SI{0}{dBm} TX power}
        \While {$P_T \leq P_T^{MAX}$} \Comment{Allowed TX power}
		\State $P_{T_i} = P_T, i\in \{1, ..., N-1\}$ \Comment{Same UAVs' TX power}
		\State Calculate $(x_0, y_0, z_0)$ \Comment{System of equations \cref{eq:gw-placement-equations-system}}
		\If{$(x_0, y_0, z_0) \neq \oslash$} \Comment{i.e., $(x_0, y_0, z_0) \in S_G$}
		\State \textbf{return} $P_T, (x_0, y_0, z_0)$ \Comment{TX power, GW UAV pos.}
		\Else
		\State $P_T = P_T + 1$ \Comment{Increase TX power by \SI{1}{\deci\belmilliwatt}}
		\EndIf
		\EndWhile
	\end{algorithmic}
\end{algorithm}
\end{tiny}

The GWP algorithm provides the same output whether downlink or uplink traffic is considered, since all the UAVs are configured with the same transmission power and the wireless channel is assumed to be symmetric. This paves the way to the usage of the GWP algorithm in emerging networking scenarios where symmetric traffic applications are growing~\cite{elshaer2014}, such as social networks, video streaming, and online gaming.\looseness=-1  

\subsection{Numerical Analysis~\label{sec:numerical_analysis} for a Simple Scenario}
Without loss of generality, we now exemplify the execution of \cref{alg:gwp} for the simple scenario shown in \cref{fig:gw-acceptance-subspace}; the algorithm is generic and may be applied to any traffic demand and number of FAPs. The scenario of \cref{fig:gw-acceptance-subspace} is composed of four FAPs that are placed at the vertices of a square of side \SI{30}{\meter}, hovering at \SI{10}{\meter} altitude. We assume the use of the IEEE 802.11ac standard with one spatial stream, \SI{800}{\nano \second} Guard Interval (GI), and \SI{160}{\mega\hertz} channel bandwidth (channel 50 at \SI{5250}{\mega\hertz}). Let us consider that the demanded capacity for the right-side FAPs is \SI{702}{Mbit/s}, which is associated to the IEEE 802.11ac MCS index 8, and the demanded capacity for the left-side FAPs is \SI{234}{Mbit/s}, which is associated to the IEEE 802.11ac MCS index 3 \cite{MCSIndex14:online}. These demanded capacity values were selected in order to exemplify the execution of \cref{alg:gwp} for an illustrative networking scenario where the righ-side FAPs have a traffic demand up to three times higher than the left-side FAPs, due to a greater number of ground users served by the first FAPs.

Taking into account the minimum SNR and the theoretical data rate of the IEEE 802.11ac MCS indexes, from \cref{tab:mapping-snr-data-rate} we conclude that the target SNR values in \SI{}{\decibel} are respectively \SI{20}{\deci\bel} for the left-side FAPs and \SI{35}{\deci\bel} for the right-side FAPs, which were calculated considering the minimum Received Signal Strength Indicator (RSSI) values proposed in \cite{MCSIndex14:online} and a \SI{-85}{dBm} noise floor power.

Solving \cref{eq:gw-placement-equations-system}, which is derived from \cref{eq:friis-propagation-model} considering $d_{max_i}$ as the Euclidean distance between $UAV_0$ and $UAV_i$, we conclude that an optimal placement for the GW UAV is $(x_0,y_0,z_0) \approx (23.3, 15.4, 3.3)$ for a transmission power $P_{T} = $  \SI{22}{\deci\belmilliwatt}. $P_T$, which is the fine-tuning parameter in \cref{eq:gw-placement-equations-system}, is initially set to \SI{0}{\deci\belmilliwatt}; then, it is iteratively increased by \SI{1}{\deci\belmilliwatt} until a valid solution for the GW UAV position is found. If no solution is found, \cref{alg:gwp} is terminated. In order to achieve a solution for the GW placement, the FAPs' positions should be adjusted, aiming at enabling shorter wireless links for a transmission power $P_T$ lower than or equal to $P_T^\textrm{MAX}$.\looseness=-1 

\begin{small}
\begin{equation} \label{eq:gw-placement-equations-system}
  \left\{
    \begin{aligned}
        & (x_0\smash{-}30)^2 + y_0^2 + (z_0\smash{-}10)^2 \leqslant \left(10^{\frac{K\smash{+}P_T\smash{-}\mymk{35}}{20}} \right)^2 \\
 	    & (x_0\smash{-}30)^2 + (y_0\smash{-}30)^2 + (z_0\smash{-}10)^2 \leqslant \left(10^{\frac{K\smash{+}P_T\smash{-}\mymk{35}}{20}} \right)^2 \\
	    & x_0^2 + (y_0\smash{-}30)^2 + (z_0\smash{-}10)^2 \leqslant \left(10^{\frac{K\smash{+}P_T\smash{-}\circled{20}}{20}} \right)^2 \\
	    & x_0^2 + y_0^2+(z_0\smash{-}10)^2 \leqslant \left(10^{\frac{K\smash{+}P_T\smash{-}\circled{20}}{20}}\right)^2 \\
	    & K \smash{=}\smash{-}20\log_{10}(5250\smash{\times}10^6) - 20\log_{10}\left(\frac{4\times\pi}{3\smash{\times}10^8}\right) - (-85) \\
    \end{aligned} 
    \right.
\end{equation}
\end{small}

In the GWP algorithm we assume that the overhead of the User Datagram Protocol (UDP), Internet Protocol (IP), and MAC packet headers is negligible; this is compliant with emerging wireless communications technologies, such as IEEE 802.11ax, where Orthogonal Frequency-Division Multiple Access (OFDMA) and frame aggregation mechanisms improve the MAC efficiency~\cite{afaqui2016}.\looseness=-1 

\begin{table}
    \setlength\abovecaptionskip{-0.1\baselineskip}
    \centering
    \caption{Data rate and minimum SNR relative to IEEE 802.11ac MCS indexes 3 and 8 (BW: \SI{160}{\mega\hertz} and GI: \SI{800}{\nano\second}).}
    \label{tab:mapping-snr-data-rate}
    \begin{tabular}{l l l}
        \hline 
        MCS index  &  MCS data rate (\si{Mbit/s}) & SNR (\SI{}{\deci\bel}) \\
        \hline 
        \qquad 3    &    \qquad\qquad 234    &      \quad\circled{20}  \\
        \qquad 8    &    \qquad\qquad 702    &      \quad\mymk{35}\\
        \hline
    \end{tabular}
\end{table}

\section{Performance Evaluation~\label{sec:performance_evaluation}}
The flying network performance achieved using the GWP algorithm is presented in this section, including the simulation setup, the simulation scenarios, and the performance metrics considered.

\subsection{Simulation Setup~\label{sec:simulation_setup}}
In order to evaluate the flying network performance achieved with the GWP algorithm, the ns-3 simulator was used. A Network Interface Card (NIC) was configured on each UAV in Ad Hoc mode, using the IEEE 802.11ac standard in channel 50, with \SI{160}{\mega\hertz} channel bandwidth, and \SI{800}{\nano\second} GI. One spatial stream was used for all inter-UAV links. The traffic generated was UDP Poisson for a constant packet size of \SI{1400}{bytes}. The data rate was automatically defined by the \emph{IdealWifiManager} mechanism. The traffic generation was only triggered after \SI{30}{\second} of simulation, in order to ensure a stable state, with a total simulation time of \SI{130}{\second}. The Controlled Delay (CoDel) algorithm~\cite{nichols2012}, which is a Linux-based queuing discipline that considers the time that packets are held in the transmission queue to discard packets, was used; it allows mitigating the bufferbloat problem. The default parameters of CoDel in ns-3 were employed~\cite{CoDel:online}.\looseness=-1 

\subsection{Simulation Scenarios~\label{sec:simulation_scenarios}}
\begin{figure}
	\centering
	\subfloat[Scenario A, in which different positions for the GW UAV were evaluated. Position 8 corresponds to the optimal GW UAV position, while position 5 corresponds to the baseline -- GW UAV placed in the FAPs center.]{
	\includegraphics[width=0.46\linewidth]{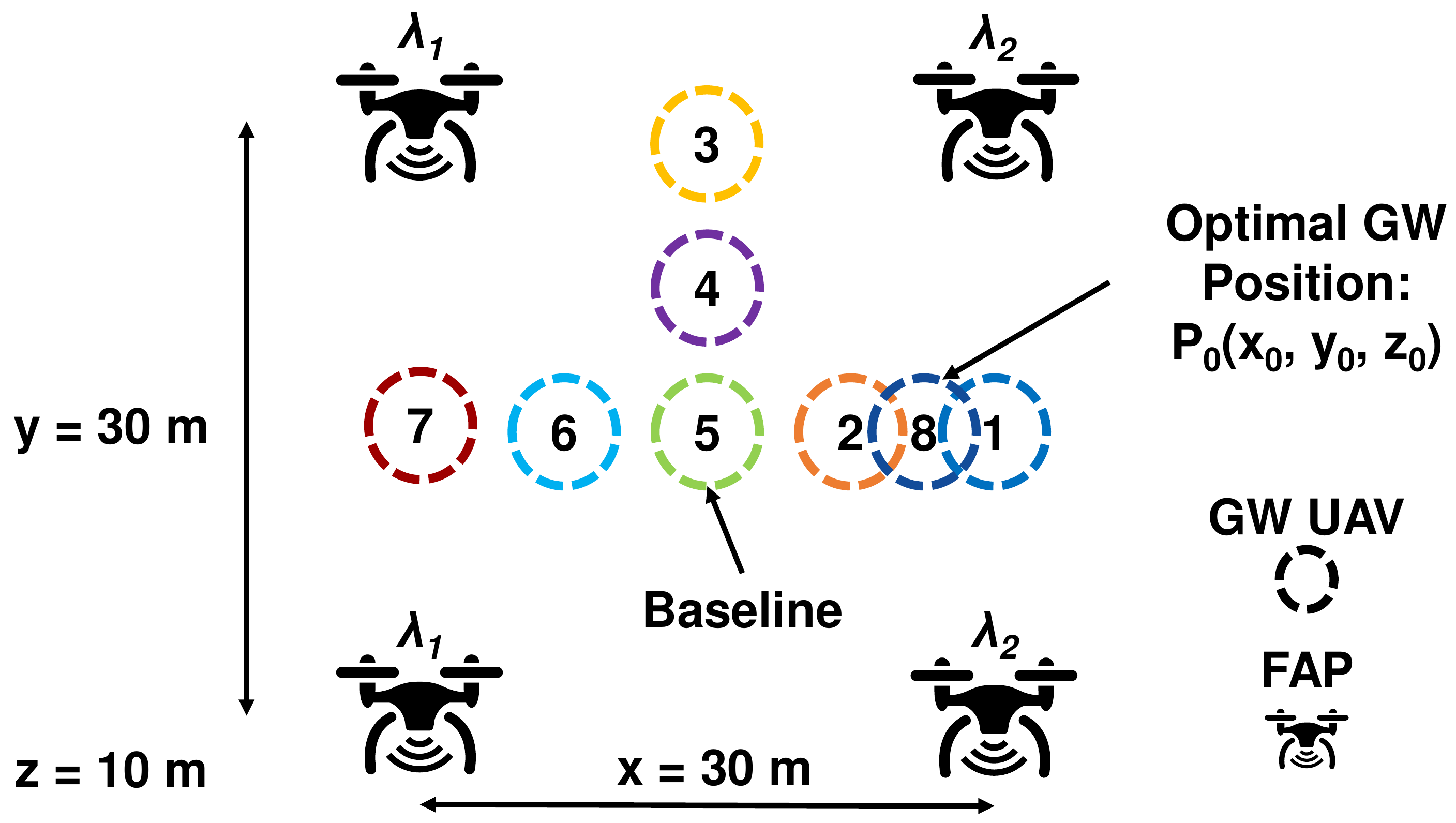}
	\label{fig:baseline-scenario}}
	\hfill
	\subfloat[Scenario B, in which 10 FAPs were randomly positioned in order to form two zones with different traffic demand: $\lambda_1$ and $\lambda_2$. The baseline, represented by a dashed circumference, corresponds to the GW UAV placed in the FAPs center.]{
		\includegraphics[width=0.44\linewidth]{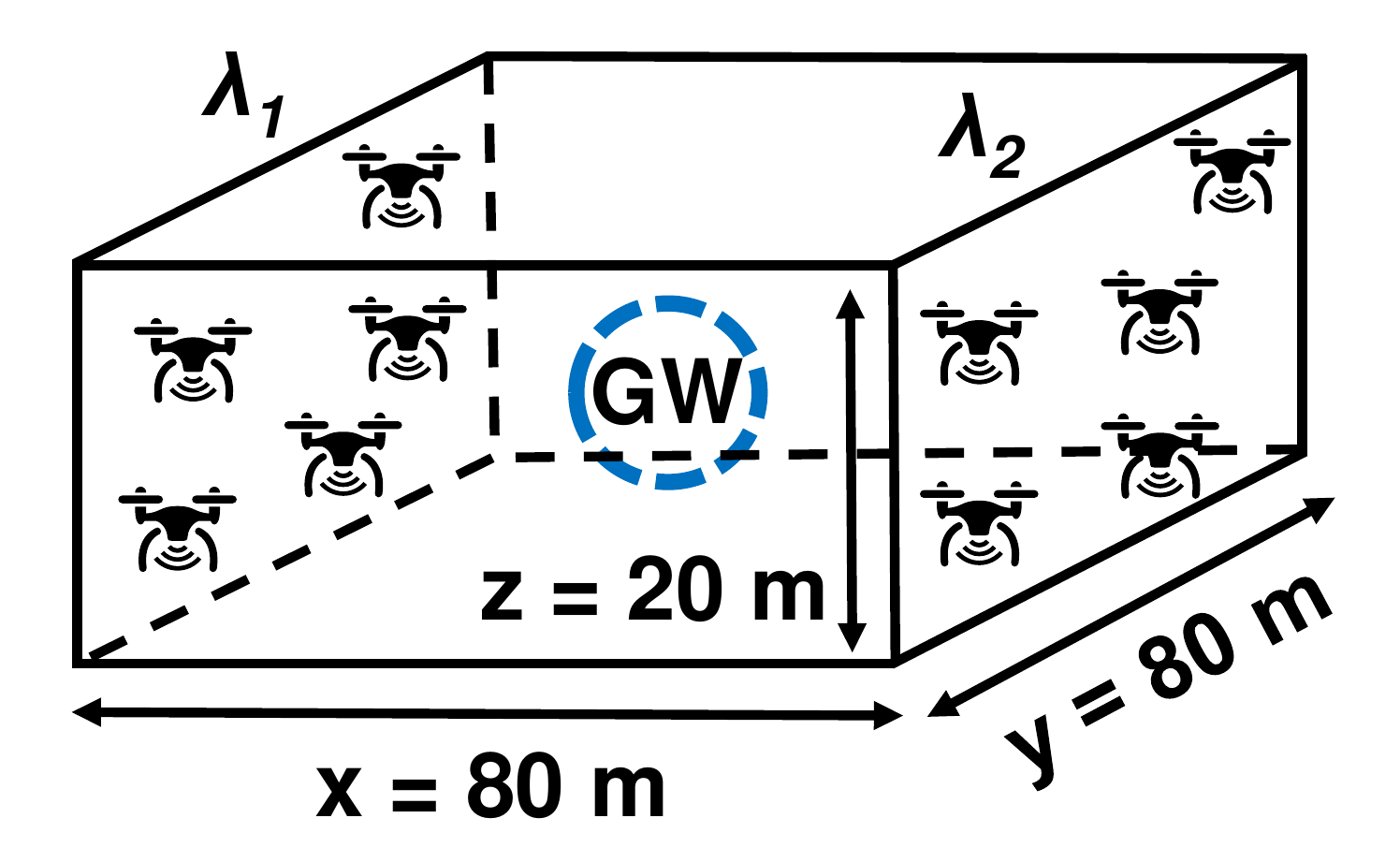}
		\label{fig:complex-scenario}}
	\hfill
	\caption{Simulation scenarios.}
	\label{PQM-Figure: Simulation scenarios}
\end{figure}

In addition to the optimal GW UAV position, which was obtained using the GWP algorithm, other positions for the GW UAV in the venue depicted in \cref{fig:baseline-scenario} were evaluated, in order to show the performance gains obtained when using the GWP algorithm; the seven additional positions considered are depicted in \cref{fig:baseline-scenario} and hereafter referred to as Scenario A. Position 1 to position 7 were defined to allow an inter-position distance of \SI{7.5}{\meter}; they aimed at exploring the vertical and horizontal corridors of the venue. We define as baseline the GW UAV placed in the FAPs center (i.e., three-coordinates average considering all FAPs), which is the position that maximizes the SNR of the wireless links between each FAP and the GW UAV; the placement of the GW UAV in the central position is commonly considered in the literature~\cite{Almeida2018}. In turn, position 8 represents the optimal GW UAV placement, which was derived from \cref{eq:gw-placement-equations-system}.\looseness=-1 

In order to evaluate the performance achieved when using the GWP algorithm in a typical crowded event, a more complex scenario, depicted in \cref{fig:complex-scenario} and hereafter named Scenario B, was also considered. It represents a flying network composed of 10 FAPs and 1 GW UAV, inside a cuboid of dimensions $\SI{80}{\meter} \times \SI{80}{\meter} \times \SI{20}{\meter}$. The FAPs were randomly positioned in order to form two zones with different traffic demand: $\lambda_1$ bit/s and $\lambda_2$ bit/s, as illustrated in \cref{fig:complex-scenario}. 

The traffic demand of the FAPs in the ns-3 simulations was defined assuming a reference fair share $L = (\SI{780}{Mbit/s}) / (N-1)$, where \SI{780}{Mbit/s} is the data rate associated to the maximum MCS index of the IEEE 802.11ac standard and $N-1$ is the number of FAPs generating traffic, considering one spatial stream, \SI{160}{\mega\hertz} channel bandwidth, and \SI{800}{\nano\second} GI. For Scenario A (cf.~\cref{fig:baseline-scenario}), we considered traffic demand $\lambda_2 = 3 \times \lambda_1 = 0.75 \times L$. For Scenario B (cf.~\cref{fig:complex-scenario}), two different traffic demand combinations were considered: i) $\lambda_2$ = $ 9 \times \lambda_1 = 0.9 \times L$, and ii) $\lambda_2$ = $ 3 \times \lambda_1 = 0.75 \times L$.

Since the GWP algorithm relies on knowing in advance the positions of the FAPs, provided by the FAPs placement algorithm in a real-world deployment, instead of generating the random waypoints during the ns-3 simulation we used BonnMotion~\cite{aschenbruck2010bonnmotion}, which is a mobility scenario generation tool. The resulting waypoints were considered to calculate in advance the optimal GW UAV position using the GWP algorithm. Finally, the optimal GW UAV position over time and the generated scenarios were imported to ns-3, with a sampling period of \SI{1}{\second}. The \textit{WaypointMobilityModel} model of ns-3 was used to place the FAPs and the GW UAV in the positions generated by BonnMotion and defined by the GWP algorithm, respectively.\looseness=-1 

The GWP algorithm was evaluated considering three counterpart approaches: 1) the GW UAV placed at the position that maximizes the SNR of the wireless links between each FAP and the GW UAV; 2) the GW UAV placed in the center of the venue; and 3) the GW UAV randomly positioned over time, following the \emph{RandomWaypointMobility} model with velocity uniformly distributed between \SI{0.5}{\meter/\second} and \SI{3}{\meter/\second}, inside the venue of \cref{fig:complex-scenario}. 

\subsection{Performance metrics~\label{sec:performance_metrics}}
The performance achieved with the GWP algorithm was evaluated considering two metrics:

\begin{itemize}
    \itemsep0em
	\item \textbf{Aggregate throughput (R):} the mean number of bits received per second by the GW UAV.
	\item \textbf{Delay:} the mean time taken by the packets to reach the sink application of the GW UAV since the instant they were generated by the source application of each FAP, including queuing, transmission, and propagation delays.
\end{itemize}

\subsection{Simulation results~\label{sec:simulation_results}}
The simulation results are presented in this section. The results were obtained after 20 simulation runs for each traffic demand combination that was considered (cf. \cref{sec:simulation_scenarios}), under the same networking conditions, using $RngSeed = 10$ and $RngRun = \{1, ..., 20\}$. The results are expressed using mean values and they are represented using the Cumulative Distribution Function (CDF) for the delay and by the complementary CDF (CCDF) for the aggregate throughput. The CCDF $F'(x)$ represents the percentage of time for which the mean aggregate throughput was higher than $x$, while the CDF $F(x)$ represents the percentage of time for which the mean delay was lower than or equal to $x$.\looseness=-1 

\begin{figure}
	\setlength\abovecaptionskip{-0.01\baselineskip}
	\centering
	\subfloat[Aggregate throughput (R) CCDF.]{
		\includegraphics[width=0.68\linewidth]{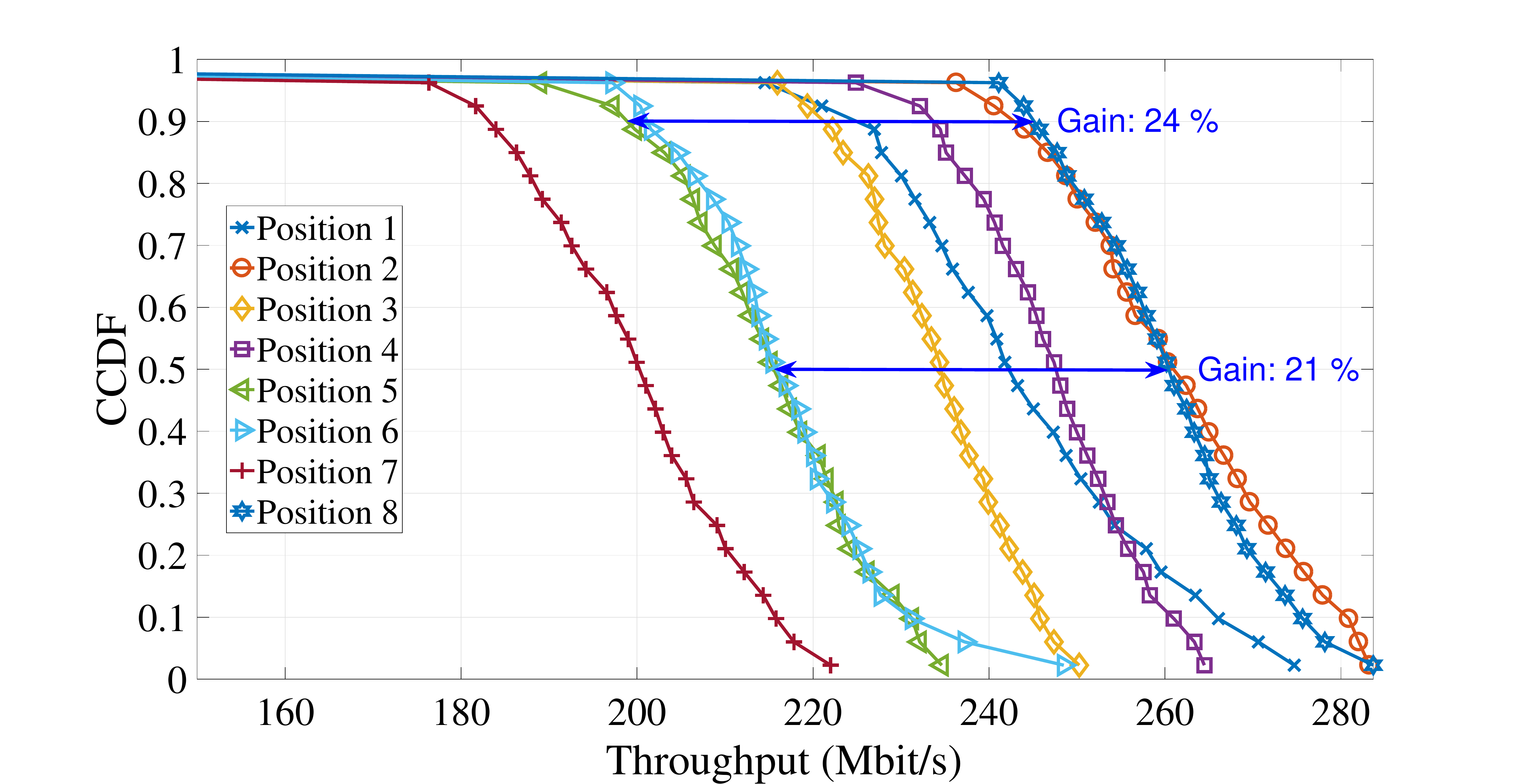}
		\label{fig:ref-throughput-ccdf}}
	\hfill
	\subfloat[Delay CDF.]{
		\includegraphics[width=0.68\linewidth]{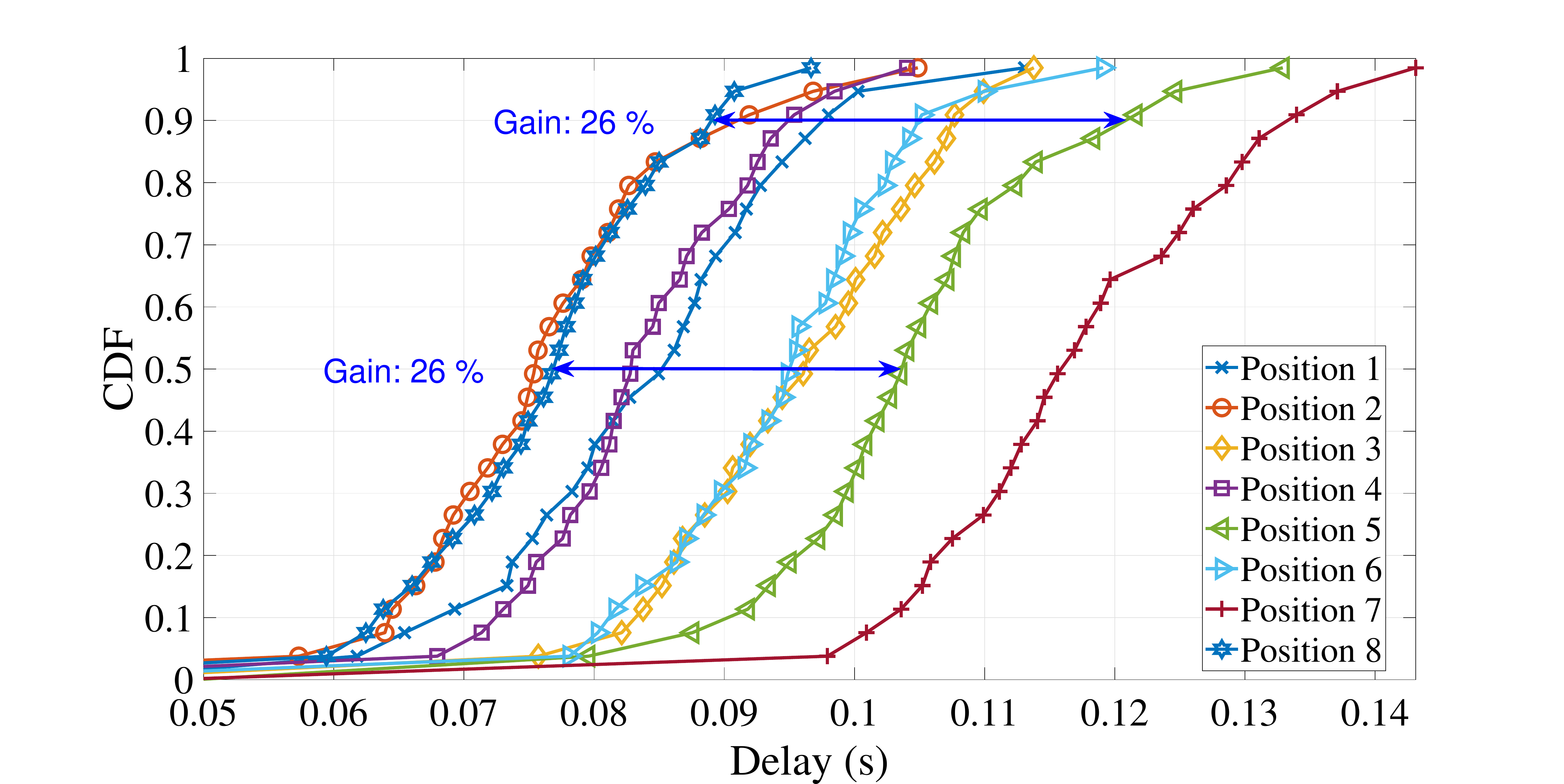}
		\label{fig:ref-delay-cdf}}
	\caption{Scenario A - Aggregate throughput (R) and delay results measured in the GW UAV. Position 8 was defined by the GWP algorithm.}
	\label{fig:ref-results}
\end{figure}

Regarding Scenario A, when the GW UAV is placed in the optimal position (Position 8 in \cref{fig:baseline-scenario}), the aggregate throughput is improved 24\% for the $90^{th}$ percentile and 21\% for the $50^{th}$ percentile (median), with respect to the baseline (i.e., the GW UAV placed in the FAPs center). In parallel, the delay is decreased 26\% for both the $90^{th}$ and $50^{th}$ percentiles (cf. \cref{fig:ref-results}). The similar results obtained for Position 2 and Position 8, which are depicted in \cref{fig:ref-results}, are justified by the closer distance between these positions; note that Position 2 was obtained by chance, while Position 8 resulted from the GWP algorithm. In order to meet the higher traffic demand of the right-side FAPs, the GWP algorithm places the GW UAV closer to them, in order to improve the SNR of the communications links and enable the selection of higher MCS indexes. This improves the overall flying network performance and the shared medium usage -- the packets are held in the transmission queues for shorter time, the transmission delay decreases, and the throughput increases.\looseness=-1

\begin{figure}
	\setlength\abovecaptionskip{-0.01\baselineskip}
	\centering
	\subfloat[Aggregate throughput (R) CCDF.]{
		\includegraphics[width=0.68\linewidth]{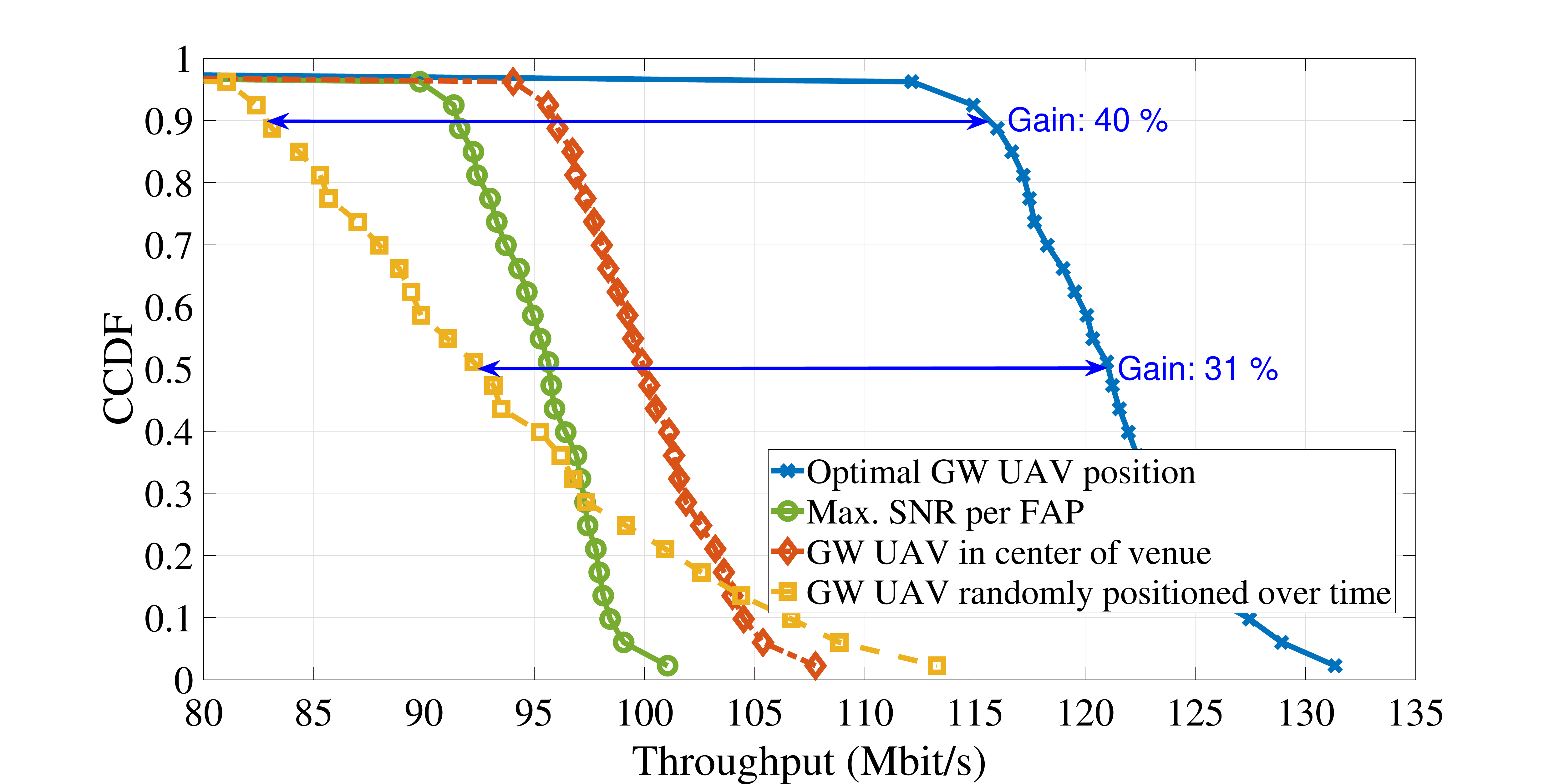}
		\label{fig:scenario10-90-throughput-ccdf}}
	\hfill
	\subfloat[Delay CDF.]{
		\includegraphics[width=0.68\linewidth]{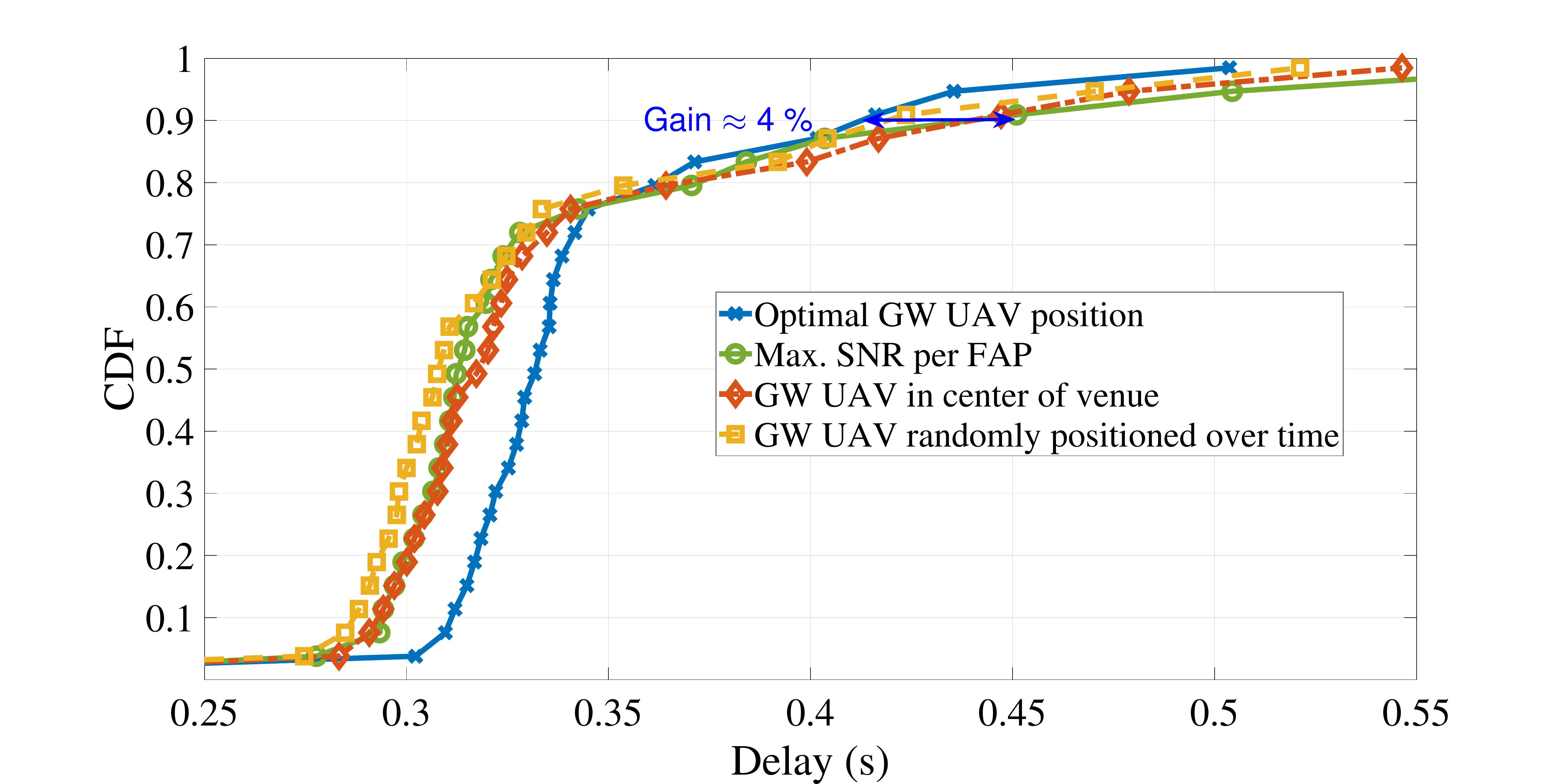}
		\label{fig:scenario10-90-delay-cdf}}
	\caption{Scenario B - Aggregate throughput (R) and delay results measured in the GW UAV for $\lambda_1 $ and $\lambda_2$ equal to 10\% and 90\% of the reference fair share $L$, respectively.}
	\label{fig:scenario10-90-results}
\end{figure}

\begin{figure}
	\setlength\abovecaptionskip{-0.01\baselineskip}
	\centering
	\subfloat[Aggregate throughput (R) CCDF.]{
		\includegraphics[width=0.68\linewidth]{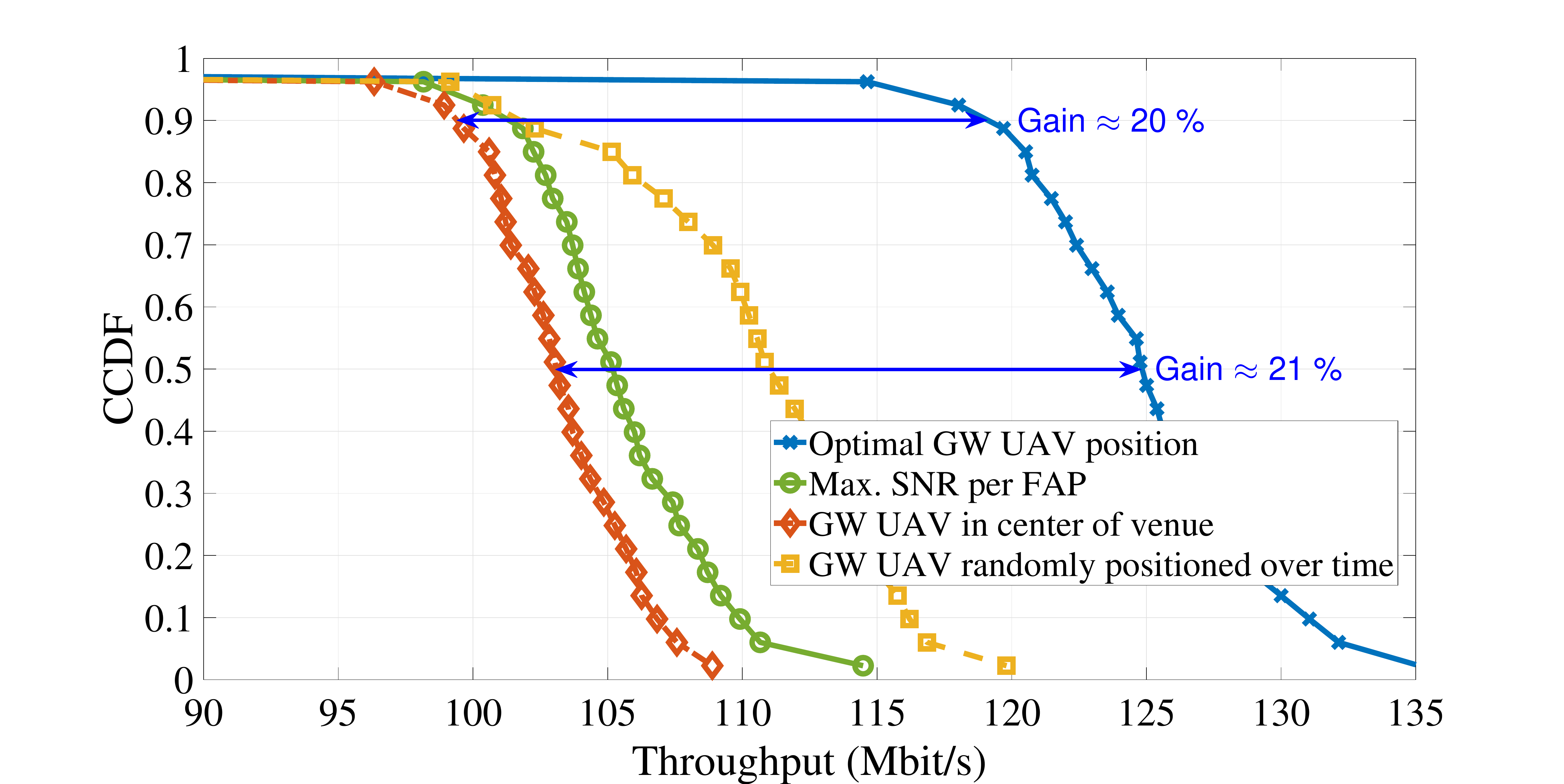}
		\label{fig:scenario25-75-throughput-ccdf}}
	\hfill
	\subfloat[Delay CDF.]{
		\includegraphics[width=0.68\linewidth]{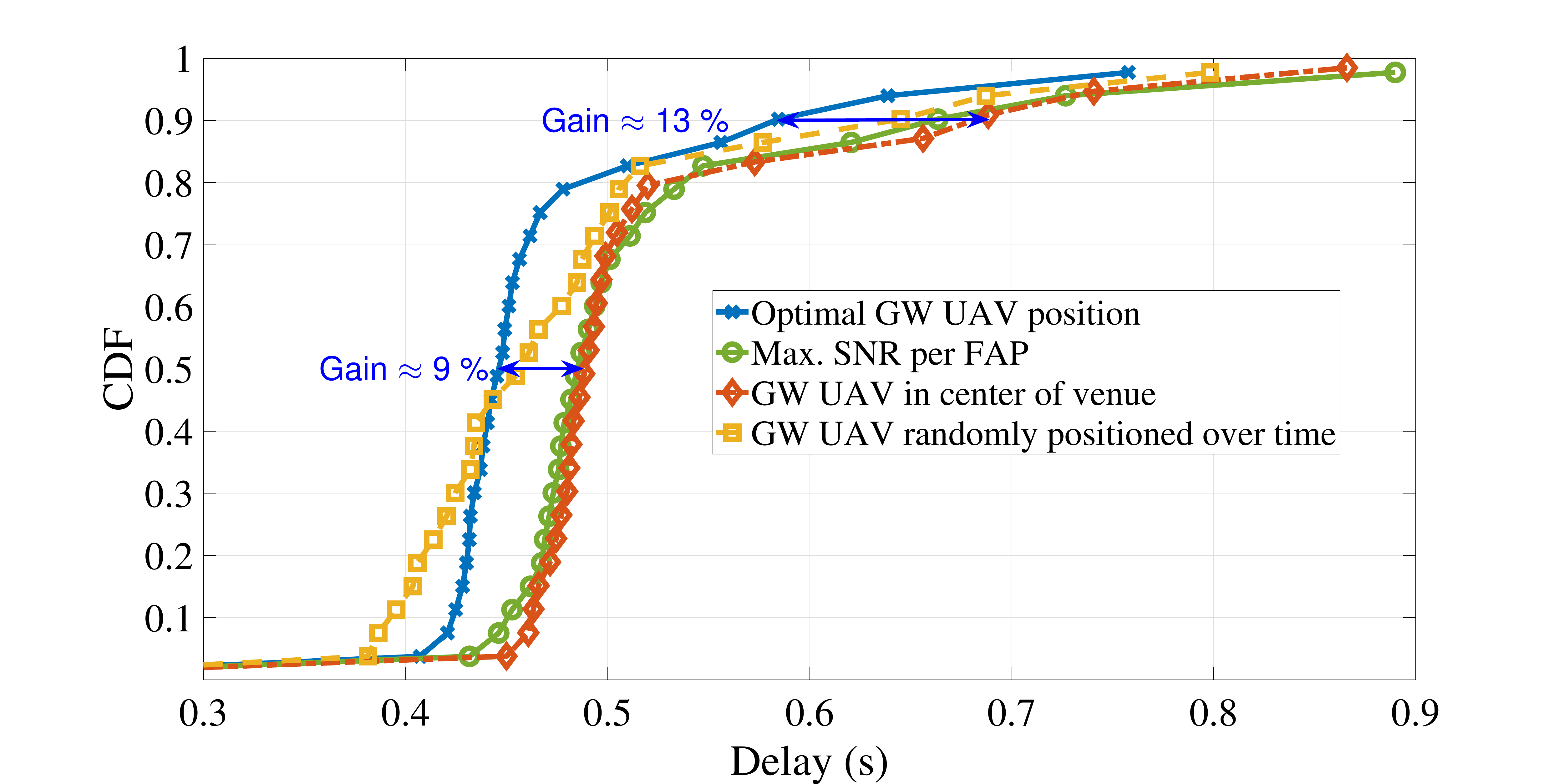}
		\label{fig:scenario25-75-delay-cdf}}
	\caption{Scenario B - Aggregate throughput (R) and delay results measured in the GW UAV for $\lambda_1 $ and $\lambda_2$ equal to 25\% and 75\% of the reference fair share $L$, respectively.}
	\label{fig:scenario25-75-results}
\end{figure}

With respect to Scenario B, when $\lambda_1 $ and $\lambda_2$ are respectively equal to 10\% and 90\% of the reference fair share $L$, the GWP algorithm allows to improve the aggregate throughput up to 40\%, considering the $90^{th}$ and $50^{th}$ percentiles, while the delay is reduced up to 4\% (cf. \cref{fig:scenario10-90-results}). When $\lambda_1 $ and $\lambda_2$ are respectively equal to 25\% and 75\% of $L$, the GWP algorithm also improves the aggregate throughput in 20\% with respect to the $90^{th}$ percentile and 21\% for the $50^{th}$ percentile; the delay is reduced 13\% for the $90^{th}$ percentile and 9\% for the $50^{th}$ percentile (cf. \cref{fig:scenario25-75-results}). These results validate the effectiveness of the GWP algorithm and corroborate our research hypothesis: the flying network performance can be improved by dynamically adjusting the position of the GW UAV over time, considering both the positions and the offered traffic of the FAPs.\looseness=-1 

\section{Conclusions~\label{sec:conclusions}}
This paper proposes GWP, a traffic-aware GW UAV placement algorithm for flying networks with controlled topology. It takes advantage of the knowledge of the offered traffic and the future topologies of the flying network, provided by a state of the art FAPs placement algorithm, to enable communications paths with high enough capacity. The flying network performance using the GWP algorithm was evaluated using ns-3 simulations. The obtained results demonstrate gains up to 40\% in aggregate throughput, while the delay is reduced up to 26\%. As future work, we aim at improving the GWP algorithm to take into account the power consumption of the GW UAV. Moreover, we will explore the traffic-aware approach proposed in this paper to position UAV relays able to forward traffic between themselves in a multi-hop flying network.\looseness=-1

\section*{Acknowledgments}
{This work is co-financed by National Funds through the Portuguese funding agency, FCT -- Fundação para a Ciência e a Tecnologia, under the PhD grant SFRH/BD/137255/2018, and by the European Regional Development Fund (FEDER), through the Regional Operational Programme of Lisbon (POR LISBOA 2020) and the Competitiveness and Internationalization Operational Programme (COMPETE 2020) of the Portugal 2020 framework [Project 5G with Nr. 024539 (POCI-01-0247-FEDER-024539)]. Part of this work was developed in the context of the course Wireless Networks and Protocols, within MAP-tele, the Portuguese Doctoral Programme in Telecommunications.}

\bibliographystyle{IEEEtran}
\bibliography{IEEEabrv,References}

\end{document}